\PassOptionsToPackage{table}{xcolor}
\documentclass[a4paper,12pt]{article}
\pdfoutput=1 % if your are submitting a pdflatex (i.e. if you have
\usepackage{jheppub} % for details on the use of the package, please
\usepackage[T1]{fontenc} % if needed

\usepackage[percent]{overpic}
\usepackage{slashed}
\usepackage{wrapfig}
\usepackage{tabu}
\usepackage{mathrsfs,amsmath,amssymb,amsthm,amsfonts,tikz,graphicx,accents,hyperref, color}
\usepackage{dsfont,epiolmec, latexsym, stmaryrd, comment}
\usepackage{slashed,ccaption}
\usepackage{mathrsfs, calligra}
\usepackage{leftidx}
\usepackage{import}
\usepackage{multirow}
\usepackage{amsfonts}
\usepackage{pifont}
\usepackage{tabularx}
\usepackage[utf8]{inputenc}
\usetikzlibrary{intersections,calc}
\usepackage{tikz-3dplot}
\usepackage{ifthen}
\usetikzlibrary{arrows}
\usepackage{amsmath}
\usepackage{xcolor}
\usepackage{geometry}
\usepackage{indentfirst}

\usepackage{array}
\DeclareFontFamily{OT1}{rsfs}{}

\DeclareFontShape{OT1}{rsfs}{m}{n}{ <-7> rsfs5 <7-10> rsfs7 <10->rsfs10}{} 

\DeclareMathAlphabet{\mycal}{OT1}{rsfs}{m}{n}

\newcommand{\di}{\text{d}}

\newcommand{\bms}{{$\mathfrak{bms}_3$}}

\newcommand{\Max}{$\mathfrak{Max}_{3}$}

\newcommand{\be}{\begin{equation}}
\newcommand{\ee}{\end{equation}}

\tdplotsetmaincoords{48}{112}
\makeatletter \@addtoreset{equation}{section}

\preprint{IPM/P-2020/017}

\title{{\boldmath {\emph{Asymptotic Symmetries of \\ Maxwell Chern-Simons Gravity with Torsion}}}}
\author[a]{ H. Adami,}
\author[b]{P. Concha,}
\author[c]{E. Rodriguez}
\author[a]{and H. R. Safari}

\affiliation{$^a$ School of Physics, Institute for Research in Fundamental
Sciences (IPM),\\ P.O.Box 19395-5531, Tehran, Iran}
\affiliation{$^b$ Departamento de Matemática y Física aplicadas, Universidad Católica de la Santísima Concepción, Alonso de Ribera 2850, Concepción, Chile}
\affiliation{$^c$ Departamento de Ciencias, Facultad de Artes Liberales, Universidad Adolfo Ibáñez, Viña del Mar-Chile.}

\emailAdd{hamed.adami@ipm.ir}

\emailAdd{patrick.concha@ucsc.cl}

\emailAdd{evelyn.rodriguez@edu.uai.cl}

\emailAdd{hrsafari@ipm.ir}

\abstract{We present a three-dimensional Chern-Simons gravity based on a deformation of the Maxwell algebra. This symmetry allows 
introduction of a non-vanishing torsion to the Maxwell Chern-Simons theory, whose action recovers the Mielke-Baelker model  for particular values of the coupling constants. By considering suitable boundary conditions, we show that the asymptotic symmetry is given by the $\widehat{\mathfrak{bms}}_3\oplus\mathfrak{vir}$ algebra with three independent central charges. 
 }

\begin{document}
\maketitle
\section{Introduction and motivations}
Three-dimensional Chern-Simons (CS) gravity theories are considered as interesting toy models since they allow us to approach diverse aspects of the gravitational interaction and underlying laws of quantum gravity. Furthermore, they share many properties with higher-dimensional gravity models which, in general, are more difficult to study. Three dimensional General Relativity (GR) with and without cosmological constant can be described through a CS action based on the AdS and Poincaré algebra, respectively \cite{Achucarro:1987vz,Witten:1988hc,Zanelli:2005sa}. Nowadays, there is a growing interest in exploring bigger symmetries in order to study more interesting and realistic physical models.

Well-known infinite-dimensional enhancements of the AdS and Poincaré symmetries, in three spacetime dimensions, are given respectively by the conformal and the $\mathfrak{bms}_{3}$ algebras. A central extension of the two-dimensional conformal algebra, which can be written as two copies of the Virasoro algebra, appears as  the asymptotic symmetry of three-dimensional GR with negative cosmological constant \cite{brown1986central}.  In the asymptotically flat case, the three-dimensional version of the Bondi-Metzner-Sachs (BMS) algebra \cite{Bondi:1962px,Sachs:1962zza, Ashtekar:1996cd,Barnich:2010eb}, denoted as $\mathfrak{bms}_{3}$, corresponds to the asymptotic symmetry of GR \cite{barnich2007classical}. $\mathfrak{bms}_{3}$ can be alternatively obtained as a flat limit of the conformal one, in a similar way as the Poincaré symmetry appears as a vanishing cosmological constant limit of AdS. The study of richer boundary dynamics could offer a better understanding of the bulk/boundary duality beyond the AdS/CFT correspondence \cite{Maldacena:1997re}. Thus, the exploration of new asymptotic symmetries of CS gravity theories based on enlarged global symmetries could be worth studying. In particular, extensions and generalizations of the conformal and the $\mathfrak{bms}_{3}$ algebras have been subsequently developed in diverse contexts in \cite{Henneaux:1999ib,Henneaux:2009pw,Skenderis:2009nt,Afshar:2011qw,Compere:2013bya,Troessaert:2013fma,Gonzalez:2014tba,Barnich:2014cwa,Fuentealba:2015jma,Fuentealba:2015wza,Perez:2016vqo,Grumiller:2016pqb,Banerjee:2016nio,Lodato:2016alv,Detournay:2016sfv,Fuentealba:2017fck,Banerjee:2017gzj,Caroca:2017onr,Concha:2018zeb,Parsa:2018kys,Concha:2018jjj,Caroca:2018obf,Banerjee:2019lrv,Caroca:2019dds}.

A particular extension and deformation of the Poincaré algebra is given by the Maxwell algebra. Such symmetry appears to describe a particle moving in a four-dimensional Minkowski background in presence of a constant electromagnetic field \cite{schrader1972maxwell,bacry1970group,Gomis:2017cmt}. This algebra is characterized by the non-vanishing commutator of the four-momentum generator $P_{a}$:
\begin{equation}
    [P_{a},P_{b}]=M_{ab}\,,
\end{equation} which is proportional to a new Abelian generator $M_{ab}$.  The Maxwell algebra and its generalizations have been useful to recover standard GR without cosmological constant from CS and Born-Infeld gravity theories in a particular limit \cite{Edelstein:2006se,Izaurieta:2009hz,Concha:2013uhq,Concha:2014vka,Concha:2014zsa}. In three spacetime dimensions, an invariant CS gravity action under Maxwell algebra has been introduced in \cite{Salgado:2014jka} and different aspects of it has been studied  \cite{Hoseinzadeh:2014bla,Concha:2015woa,Caroca:2017izc,Aviles:2018jzw,Bansal:2018qyz,Concha:2018jxx, Concha:2019icz,Concha:2019mxx,Chernyavsky:2020fqs,Concha:2020sjt}. As Poincar\'e symmetry, the Maxwell symmetry describes a three-dimensional gravity theory whose geometry is Riemannian and locally flat. However, the presence of an additional gauge field in the Maxwell case leads to new effects compared to GR. In particular, in \cite{Concha:2018zeb}, the authors have shown that the Maxwellian gravitational gauge field modifies not only the vacuum energy and angular momentum of the stationary configuration but also the asymptotic structure.

To accommodate a non-vanishing torsion to the Maxwell CS gravity theory it is necessary to deform the Maxwell algebra. Here, we show that a particular deformation of the Maxwell symmetry, which we refer to as ``deformed Maxwell algebra'', allows us to introduce not only a torsional but also a cosmological constant term along the Einstein-Hilbert term.  Then, motivated by the recent results on the Maxwell algebra, we explore the effects of deforming the Maxwell symmetry both to the bulk and boundary dynamics. At the bulk level, we show that the invariant CS gravity action under the deformed Maxwell algebra reproduces the Maxwell field equations but with a non-vanishing torsion describing a Riemann-Cartan geometry. Interestingly, the CS action can be seen as a Maxwell version of a particular case of the Mielke-Baekler (MB) gravity theory \cite{Mielke:1991nn} which describes a three-dimensional gravity model in presence of non-vanishing torsion. Further studies of the MB gravity have been subsequently developed in \cite{Baekler:1992ab, Blagojevic:2003vn,Blagojevic:2003wn,Cacciatori:2005wz,Blagojevic:2006jk,Blagojevic:2006hh,Giacomini:2006dr,Klemm:2007yu,Santamaria:2011cz,Cvetkovic:2018ati,Peleteiro:2020ubv}. Here we explore the effects of having a non-vanishing torsion in Maxwell CS gravity at the level of the boundary dynamics. In particular, by considering suitable boundary conditions, we show that the asymptotic symmetry can be written as the $\widehat{\mathfrak{bms}}_3\oplus \mathfrak{vir}$ algebra. This infinite-dimensional symmetry was recently obtained as a deformation of the infinite-dimensional enhancement of the Maxwell algebra, denoted as $\mathfrak{Max}_{3}$ algebra \cite{Concha:2019eip}. We also show that the vanishing cosmological constant limit $\ell\rightarrow\infty$ can be applied not only at the CS gravity theory level but also at the asymptotic algebra, leading  to the Maxwell CS gravity and its respective asymptotic symmetry previously introduced in \cite{Concha:2018zeb}. 

The paper is organized as follows. In section \ref{sec:2}, we present the three-dimensional CS gravity theory which is invariant under a particular deformation of the Maxwell algebra. Furthermore, considering asymptotically flat geometries with null boundary, we discuss the BMS-like solution of the theory. We provide boundary conditions allowing a well-defined action principle. In section \ref{sec:3}, we show that the asymptotic symmetry algebra for the Maxwell CS gravity with torsion is given by an infinite enhancement of a deformed Maxwell algebra,  which  can be written as the direct sum $\widehat{\mathfrak{bms}}_3\oplus \mathfrak{vir}$. Finally, in section \ref{sec-discussions} we discuss the obtained results and possible future developments.

\paragraph{Notation.} We adopt the same notation as \cite{Parsa:2018kys, Safari:2019zmc,Concha:2019eip} for the algebras; for algebras we generically use ``mathfrak'' fonts, like $\mathfrak{vir}$, \bms\ and $\mathfrak{Max}_{3}$. 
The centrally extended version of an algebra $\mathfrak{g}$ will be denoted by $\hat{\mathfrak{g}}$, e.g. Virasoro algebra $\mathfrak{vir}=\widehat{\mathfrak{witt}}$. 
%%%%%%%%%%%%%%%%%%%%%%%%%%%%%%%%%%%%%%%%%%%%%%%%%%%%%%%%%%%%%%%%%%%%%%%
%%%%%%%%%%%%%%%%%%%%%%%%%%%%%%%%%%%%%%%%%%%%%%%%%%%%%%%%%%%%%%%%%%%%%%%%%%%%

\section{Maxwell Chern-Simons gravity theory with torsion}\label{sec:2}

Using the CS formalism, we present the three-dimensional gravity theory based on a particular deformation of the Maxwell algebra. Unlike the Maxwell case, such deformation leads to a non-vanishing torsion as equation of motion. The deformed Maxwell algebra is spanned by the generators $\{J_a,P_a,M_a\}$, which satisfy the following non-vanishing commutation relations:
%The Einstein-Hilbert action in the Palatini formulation is equivalent (up to boundary terms) to a Chern-Simons action \cite{Witten:1988hc}. Let us start with the Chern-Simons theory in $2+1$ dimensions with gauge group $\textbf{G}$ (generated by Lie algebra $\mathfrak{g}$) on a manifold with the topology $\mathcal{M} = \mathbb{R} \times \Sigma$ where $\Sigma$ is a two-dimensional manifold with boundary $\partial \Sigma \cong S^{1}$. Let $0 \leq r < \infty$ denote radial direction and retarded time coordinate $-\infty < u <\infty$ as well as angular coordinate $\phi \sim \phi + 2 \pi$ parameterizes the boundary of the cylinder. The Chern-Simons action is given by
%\begin{equation}\label{CS-Action}
    %S_{\text{CS}}[A]= \frac{k}{4 \pi} \int_{\mathcal{M}} \langle A \wedge d A+ \frac{2}{3} A \wedge A \wedge A  \rangle
%\end{equation}
%where $A$ is the gauge connection, $\langle \, , \, \rangle$ denotes the invariant trace, and $k=1/(4G)$ is Chern-Simons level. The connection one-form takes values in certain Lie algebras. In this work, we consider a particular deformation of the Maxwell algebra which has been introduced in \cite{Concha:2019eip}. Such deformation is spanned by the set of generators $\{J_a,P_a,M_a\}$ which satisfy the following non-vanishing commutation relations:
\begin{equation}\label{algebra01} 
\begin{split}
 & [J_a,J_b]=\epsilon_{ab}^{\hspace{3 mm}c} J_{c} \,, \\
 &[J_a,P_b]=\epsilon^{\hspace{3 mm}c}_{ab}P_{c} \,,\\
 &[J_a,M_b]=\epsilon^{\hspace{3 mm}c}_{ab} M_{c} \,,\\
 &[P_a,P_b]=\epsilon^{\hspace{3 mm}c}_{ab} (M_{c}+\frac{1}{\ell}P_{c}) \,,
\end{split}
\end{equation}
where $\epsilon_{abc}$ is the three-dimensional Levi-Civita tensor and $a,b=0,1,2$ are the Lorentz indices which are lowered and raised with the Minkowski metric $\eta_{ab}$. The $\ell$ parameter appearing in the last commutator is related to the cosmological constant $\Lambda$. Then, the vanishing cosmological constant limit $\ell\rightarrow\infty$ reproduces the Maxwell symmetry. Let us note that the Hietarinta-Maxwell algebra \cite{Hietarinta:1975fu,Bansal:2018qyz,Chernyavsky:2020fqs} is recovered in the limit $\ell\rightarrow\infty$ when the role of the $P_{a}$ and $M_{a}$ generators is interchanged. One can see that $J_a$ and $P_a$ are not the generators of a Poincaré subalgebra. However, as it is pointed out
in \cite{Concha:2019eip}, \eqref{algebra01} can be rewritten as the $\mathfrak{iso}(2,1)\oplus\mathfrak{so}(2,1)$ algebra. This can be seen by a redefinition of the generators,
%algebra considering the redefinition of the generators as follows:
%\begin{equation} 
%\begin{split}\label{redefine-pp}
 %& J_a\equiv L_{a}+S_{a} \,,\\
 %& P_a\equiv \left(T_{a}+\frac{1}{\ell}S_{a}\right) \,,\\
 %& M_a\equiv -\frac{1}{\ell}T_{a} \,,
%\end{split}
%\end{equation}

\begin{equation} 
\begin{split}\label{redefine-pp}
 & L_{a}\equiv J_{a} - \ell P_{a}- \ell^2 M_{a} \,,\\
 & S_a\equiv \ell P_{a}+ \ell^2 M_{a} \,,\\
 & T_a\equiv - \ell \,M_{a} \,,
\end{split}
\end{equation}
where $L_{a}$ and $T_{a}$ are the respective generators of the $\mathfrak{iso}(2,1)$ algebra, while $S_{a}$ is a $\mathfrak{so}(2,1)$ generator. Then, the Lie algebra \eqref{algebra01} can be rewritten as
%\begin{equation} 
%\begin{split}
 %& [\Tilde{J}_a,\Tilde{J}_b]=\epsilon_{ab}^{\hspace{3 mm}c} \Tilde{J}_{c} \,, \\
 %& [\Tilde{J}_a,\Tilde{P}_b]=\epsilon_{ab}^{\hspace{3 mm}c} \Tilde{P}_{c} \,, \\
% &[\Tilde{M}_a,\Tilde{M}_b]=\epsilon^{\hspace{3 mm}c}_{ab} \Tilde{M}_{c} \,.
%\end{split}
%\end{equation}
\begin{equation} 
\begin{split}
 & [L_a,L_b]=\epsilon_{ab}^{\hspace{3 mm}c} L_{c} \,, \\
 & [L_a,T_b]=\epsilon_{ab}^{\hspace{3 mm}c} T_{c} \,, \\
 &[S_a,S_b]=\epsilon^{\hspace{3 mm}c}_{ab} S_{c} \,.
\end{split}
\end{equation}

It is important to point out that \eqref{algebra01} is not the unique way of deforming the Maxwell algebra. 
It is shown in \cite{Gomis:2009dm} that
%As was shown in \cite{Gomis:2009dm},
the Maxwell algebra can be deformed into two different algebras: $\mathfrak{so}(2,2)\oplus\mathfrak{so}(2,1)$ and $\mathfrak{iso}(2,1)\oplus\mathfrak{so}(2,1)$. The former has been largely studied in \cite{Diaz:2012zza,Hoseinzadeh:2014bla,Concha:2018jjj,Concha:2019lhn} whose asymptotic symmetry is described by three copies of the Virasoro algebra \cite{Caroca:2018obf,Concha:2018jjj}, while the latter has only been approached through a deformation process \cite{Gomis:2009dm,Concha:2019eip}. In the present work, using the basis $\left\{ J_a,P_a,M_a\right\}$, we find asymptotic symmetry of the CS gravity theory based on the $\mathfrak{iso}(2,1)\oplus\mathfrak{so}(2,1)$ algebra.
%we shall focus on finding the asymptotic symmetry of the CS gravity theory based on the $\mathfrak{iso}(2,1)\oplus\mathfrak{so}(2,1)$ algebra, but using the basis $\left\{ J_a,P_a,M_a\right\}$ satisfying \eqref{algebra01}.
The motivation to use such basis is twofold. First, it allows us to recover the Maxwell CS gravity theory in a particular limit. Second, as we shall see, it reproduces the Maxwell field equations with a non-vanishing torsion.

A three-dimensional gravity can be formulated as a CS theory described by the action
%A three-dimensional gravity action based on a given Lie algebra can be formulated as a CS form on a manifold $\mathcal{M}$,
\begin{equation}\label{CS-Action}
    S_{\text{CS}}[A]= \frac{k}{4 \pi} \int_{\mathcal{M}} \langle A \di A+ \frac{2}{3} A^{3}  \rangle \,,
\end{equation}
with a given Lie algebra on a manifold $\mathcal{M}$, where $A$ is the gauge connection, $\langle \, , \, \rangle$ denotes the invariant trace and $k=1/(4G)$ is the CS level. For the sake of simplicity we have omitted writing the wedge product. 
%\hanote{I think, it is convenient to use a 3D vector algebra notation for the Lorentz vectors in which
%contractions with $\eta_{a b}$ and $\epsilon_{abc}$ are denoted by dots and crosses, respectively. I do not insest to do so but if you agree please let me know. Then I can change the notation.}
%\enote{Personally I prefer notation we have used. I like to explicitly write the Lorentz indices and the corresponding contractions. But it is my personal opinion. Why do you think it is convenient to introduced another notation?}
The gauge connection one-form $A$ for the deformed Maxwell algebra reads
\begin{equation}\label{one-form}
    A= e^{a} P_{a}+\omega^{a}J_{a}+f^{a} M_{a} \,,
\end{equation}
where $e^{a}$, $\omega^{a}$ and $f^{a}$ are the dreibein, the (dualized) spin connection and an auxiliary one-form field, respectively. %The corresponding two-form curvature  $F=dA+\frac{1}{2}[A,A]$ is given by
The associated field strength $F=\di A+\frac{1}{2}[A,A]$ can be written as
\begin{equation}\label{two-form}
    F= K^{a} P_{a}+R^{a}J_{a}+W^{a} M_{a},
    %F= \hat{T}^{a} P_{a}+R^{a}J_{a}+F^{a} M_{a},
\end{equation}
where
\begin{subequations}
\begin{align}\label{curvatures}
   & R^{a}=\di \omega^{a}+\frac{1}{2}\epsilon^a_{ \, \, \,  bc}\omega^{b}\omega^{c} \,,\\
   & K^{a}= T^{a}+\frac{1}{2\ell}\epsilon^a_{ \, \, \,  bc} e^{b}e^{c} \,, \\ 
   & W^{a}=D(\omega) f^{a} +\frac{1}{2}\epsilon^a_{ \, \, \,  bc} e^{b}e^{c} \,.
    %&R_{a}=d\omega_{a}+\frac{1}{2}\epsilon_{abc}\omega^{b}\omega^{c} \,,\\
    %&\hat{T}_{a}= T_{a}+\frac{1}{2\ell}\epsilon_{abc} e^{b}e^{c} \,, \\ 
    %&F=D_{\omega}f_{a} +\frac{1}{2}\epsilon_{abc} e^{b}e^{c} \,.
    \end{align}
\end{subequations}
Here, $T^ {a}=D(\omega) e^{a}$ is torsion two-form, $R^{a}$ is curvature two-form, and $D(\omega)\Phi^{a}=\di \Phi^{a}+\epsilon^a_{ \, \, \,  bc}\omega^{b}\Phi^{c}$ is the exterior covariant derivative. Naturally, the flat limit $\ell\rightarrow\infty$ reproduces the Maxwell field strength \cite{Concha:2018zeb}. On the other hand, the non-degenerate bilinear form of the algebra \eqref{algebra01} reads
\begin{equation}
\begin{split}\label{invtensor}
     &\langle J_{a} J_{b} \rangle = \alpha_0 \eta_{a b}, \hspace{0.7 cm} \langle P_{a} P_{b} \rangle = (\frac{\alpha_1}{\ell} + \alpha_2) \eta_{a b},  \\
     & \langle J_{a} P_{b} \rangle = \alpha_1 \eta_{a b}, \hspace{0.7 cm} \langle P_{a} M_{b} \rangle = 0,\\
     &\langle J_{a} M_{b} \rangle = \alpha_2 \eta_{a b}, \hspace{0.7 cm} \langle M_{a} M_{b} \rangle = 0,
 \end{split}  
\end{equation}
where $\alpha_0, \alpha_1$ and $\alpha_2$ are arbitrary constants. One can see that the flat limit $\ell\rightarrow\infty$ leads to the non-vanishing components of the invariant tensor for the Maxwell algebra \cite{Concha:2018zeb}.

Considering the one-form gauge potential \eqref{one-form} and the non-vanishing components of the invariant tensor \eqref{invtensor}, one can rewrite the CS action \eqref{CS-Action} as
%A CS action invariant under the algebra \eqref{algebra01} can be written considering the one-form gauge potential \eqref{one-form} and the non-vanishing components of the invariant tensor \eqref{invtensor} in the general definition of the CS action \eqref{CS-Action},
\begin{equation}
\begin{split}\label{CS}
    S_{CS} = \frac{k}{4 \pi} \int_{\mathcal{M}} \bigg\{ &\alpha_0 \left( \omega^a \di \omega_a + \frac{1}{3}\epsilon^{abc}\omega_a \omega_b \omega_c
    \right) + \alpha_1 \left( 2 R_a e^a + \frac{1}{3\ell^{2}} \epsilon^{abc} e_a e_b e_c + \frac{1}{\ell}T^a e_a \right) \\
    &+ \alpha_2 \left( T^a e_a +2R^a f_a + \frac{1}{3\ell} \epsilon^{abc} e_a e_b e_c \right) \bigg\}
\end{split}
\end{equation}
%S = \frac{k}{4 \pi} \int_{\mathcal{M}} \bigg\{ &-\sigma e \cdot R(\omega)+\frac{\Lambda_{0}}{6} e \cdot e \times e +\frac{1}{2 \mu} \left( \omega \cdot d \omega + \frac{1}{3} \omega \cdot \omega \times \omega\right) \\
  % &+ \frac{\Lambda_{0}}{2}  e \cdot T(\omega) +\beta f \cdot R(\omega) \bigg\}
up to a surface term. One can see that the CS action is proportional to three independent sectors each one with its respective coupling constant $\alpha_i$. In particular, the first term is the so-called exotic Lagrangian \cite{Witten:1988hc}. The second term contains the usual Einstein Lagrangian with cosmological constant term plus a torsional term related to the so-called Nieh-Yan invariant density.

It is interesting to notice that the CS gravity action \eqref{CS} can be seen as a Maxwell extension of a particular case of the MB model, which describes a three-dimensional gravity model in presence of non-vanishing torsion. The MB action is given by
\begin{equation}\label{MB}
    I_{MB}=aI_{1}+\Lambda I_{2}+\beta_{3} I_{3}+\beta_{4} I_{4}
\end{equation}
where $a,\Lambda,\beta_{3}$ and $\beta_{4}$ are constants and
\begin{eqnarray}
I_{1} & = & 2\int e_{a}R^{a}\,,\nonumber\\
I_{2} & = & -\frac{1}{3}\int \epsilon_{abc}e^{a}e^{b}e^{c}\,,\nonumber\\
I_{3} & = & \int \omega^a \di \omega_a + \frac{1}{3}\epsilon^{abc}\omega_a \omega_b \omega_c\,,\label{MBterms}\\
I_{4} & = &\int e_{a}T^{a}\,.\nonumber
\end{eqnarray}
Particularly, in the absence of the auxiliary field $f^a$ in \eqref{CS}, the constants appearing in the MB gravity can be identified with those of the   the deformed Maxwell algebra CS theory as
\begin{equation}\label{MBcts}
    a=\frac{k}{4\pi}\alpha_{1}\,, \qquad \Lambda=-\frac{k}{4\pi \ell}\left( \frac{\alpha_{1}}{\ell}+\alpha_2\right)\,, \qquad \beta_{3}=\frac{k}{4\pi}\alpha_{0}\, \qquad \beta_{4}=\frac{k}{4\pi }\left( \frac{\alpha_{1}}{\ell}+\alpha_2\right)\,.
\end{equation}
Thus, the CS gravity action \eqref{CS} can be interpreted as the Maxwellian version of a particular case of the MB gravity action when the MB's constants satisfy \eqref{MBcts}.

It is important to point out that one can accommodate a generalized cosmological constant in the Maxwell gravity theory using another deformation of the Maxwell algebra, known as AdS-Lorentz algebra \cite{Soroka:2004fj}. However, in the AdS-Lorentz case, besides the Einstein-Hilbert term there is an additional gauge field $f_a$, while there is no  torsion term $\alpha_{1}$ \cite{Diaz:2012zza,Hoseinzadeh:2014bla,Concha:2018jjj,Concha:2019lhn}. In the present case, the deformed Maxwell symmetry allows us to introduce both a cosmological constant and  a torsion term. In the flat limit $\ell\rightarrow\infty$ the CS action reproduces the Maxwell CS gravity action which contains pure GR as sub-case.  Dynamics of the $f_a$ gauge field is completely determined by the last term with coupling constant $\alpha_2$. In particular, the equations of motion appear by considering the variation of the action \eqref{CS} under the respective gauge fields:
\begin{eqnarray}
\delta e^{a} & : & \qquad0=\alpha_{1}\left(2R^{a}+\frac{2}{\ell}K^{a}\right)+\alpha_{2}K^{a},\nonumber \\
\delta\omega^{a} & : & \qquad0=\alpha_{0}R^{a}+\alpha_{1}K^{a}+\alpha_{2}W^{a},\label{eom}\\
\delta f^{a} & : & \qquad0=\alpha_{2}R^{a},\nonumber
\end{eqnarray}
%\begin{eqnarray}
%\delta e^{a} & : & \qquad0=\alpha_{1}\left(2R_{a}+\frac{2}{\ell}\hat{T}_{a}\right)+\alpha_{2}\hat{T}_{a},\nonumber \\
%\delta\omega^{a} & : & \qquad0=\alpha_{0}R_{a}+\alpha_{1}\hat{T}_{a}+\alpha_{2}F_{a},\label{eom}\\
%\delta f^{a} & : & \qquad0=\alpha_{2}R_{a},\nonumber
%\end{eqnarray}
Then, when $\alpha_2\neq 0$ we find the curvature two-forms \eqref{two-form} should vanish,
\begin{equation}\label{EOM}
    R^{a}= 0\,, \qquad 
    K^{a}=0\,, \qquad
    W^{a}=0 \, .
\end{equation}
%\begin{equation}
%\begin{split}\label{EOM}
%    &R_{a}= 0\,, \\ 
%    &\hat{T}_{a}=0\,, \\
%    &F_{a}=0\,.
%    \end{split}
%\end{equation}
Note that the CS gravity theory \eqref{CS} describes the Maxwell CS gravity theory in presence of a non-vanishing torsion $T^{a}\neq 0$. In particular, the first two equations $R^{a}=0$ and $T^{a}=-\frac{1}{2 \ell}\epsilon^{a}_{\,\,\, bc}e^{b}e^{c}$ correspond to the three-dimensional teleparallel theory in which the cosmological constant can be seen as a source for the torsion.  On the other hand, the vanishing of $W^a$ implies
%$F_{a}$curvature 
that the exterior covariant derivative of the auxiliary field $f_a$
is constant. In particular, in the flat limit $\ell\rightarrow\infty$ the field equation for $f_{a}$ remains untouched and is analogue to the constancy of the electromagnetic field in flat spacetime.

One can see that each term of the action \eqref{CS} is invariant under the gauge transformation laws of the algebra \eqref{algebra01}. Indeed, considering
\begin{equation}
    \Lambda=\varepsilon^{a}P_{a}+\rho^{a}J_{a}+\chi^{a}M_{a},
\end{equation}
we have that the gauge transformations $\delta A =\di \Lambda +[A,\Lambda]$ of the theory are given by
\begin{equation}
   \begin{split}
        &\delta_{\Lambda} e^{a}=D(\omega)\varepsilon^{a}-\epsilon^{abc}\rho_{b}e_{c}+\frac{1}{\ell}\epsilon^{abc}e_{b}\varepsilon_{c}, \\
        &\delta_{\Lambda}\omega^{a}=D(\omega)\rho^{a}, \\
        &\delta_{\Lambda}f^{a}=D(\omega)\chi^{a}+\epsilon^{abc}e_{b}\varepsilon_{c}-\epsilon^{abc}\rho_{b}f_{c}.
    \end{split}
\end{equation}

In this work, we  analyze the consequences of this particular deformation of the Maxwell symmetry at the level of the %conserved charges and
asymptotic structure. In the Maxwell case, as was shown in \cite{Concha:2018zeb}, the presence of the additional gauge field $f^{a}$ leads to new effects compared to GR and the asymptotic symmetries is found to be a deformed $\mathfrak{bms}_{3}$,  denoted as $\mathfrak{Max}_{3}$ in \cite{Concha:2019eip}. Here, we explore the implications of deforming the Maxwell algebra as in \eqref{algebra01}.
        %&\delta_{\Lambda}\omega^{a}=D_{\omega}\rho^{a}, \\
       % &\delta_{\Lambda}\f^{a}=D_{\omega}\xi^{a}+\epsilon^{abc}e_{b}\varepsilon_{c}-\epsilon^{abc}\rho_{b}f_{c}.
%%%%%%%%%%%%%%%%%%%%%%%%%%%%%%%%%%%%%%%%%%%%%%%%%%%%%%%%%%%%%%%%%%%%%%%%%%%
%%%%%%%%%%%%%%%%%%%%%%%%%%%%%%%%%%%%%%%%%%%%%%%%%%%%%%%%%%%%%%%%%%%%%%%%%%%%

%%%%%%%%%%%%%%%%%%%%%%%%%%%%%%%%%%%%%%%%%%%%%%%%%%%%%%%%%%%%%%%%%%%%%%%%%%%%%%
%%%%%%%%%%%%%%%%%%%%%%%%%%%%%%%%%%%%%%%%%%%%%%%%%%%%%%%%%%%%%%%%%%%%%%%%%%%%

Let us recall that given an action there are two ways to render it having a well-posed variation principle. One of them is to add boundary terms to the action and the other  is imposing suitable boundary conditions on fields. Let us consider the variation of the action \eqref{CS-Action},
\begin{equation}
    \delta S_{\text{CS}}[A]= \frac{k}{2 \pi} \int_{\mathcal{M}} \langle \delta A F  \rangle  + \frac{k}{4 \pi} \int_{\partial\mathcal{M}} \langle \delta A\, A \rangle \,,
\end{equation}
%where
%\begin{equation}\label{field-strength}
%    F = d A + \frac{1}{2}[A, A]\,,
%\end{equation}
%is the field strength.
where $\partial\mathcal{M}$ is the boundary of $\mathcal{M}$. The field equations require vanishing of the field strength and the on-shell boundary contribution to the action is the surface term
\begin{equation}\label{bdy}
    \delta S_{\text{CS}}[A]\big|_{\text{bdy}} = - \frac{k}{4 \pi} \int_{\partial\mathcal{M}} \langle  A  \delta A \rangle\, .
\end{equation} 
In our case this term reads as
\begin{equation}
    \delta S_{\text{CS}}\big|_{\text{bdy}} = \frac{k}{4 \pi} \int_{\partial\mathcal{M}}  \left[ \delta\omega^{a}\left(\alpha_{0}\omega_{a}+\alpha_{1} e_{a}+\alpha_{2}f_{a}\right)+\delta e^{a}\left(\alpha_{1}\omega_{a}+\left(\frac{\alpha_1}{l}+\alpha_{2}\right)e_{a}\right)+\alpha_{2}\delta f^{a}\omega_{a}\right] \, .
\end{equation}
%As we can see from the above expression, it depends on the boundary $\partial\mathcal{M}$ we are considering.
We will see later that for spacetimes with null boundary, where the boundary is located at $r=const\rightarrow\infty$, the action principle is satisfied without addition of boundary terms.

\subsection{BMS-like solution}
In this section we analyze the field equations (\ref{EOM}). We consider spacetimes with null boundary, which can be described in the three-dimensional BMS gauge.   We parametrize spacetime by the local coordinates $x^{\mu}=\left(u,r,\phi\right)$, where $-\infty < u <\infty$ is the retarded time coordinate,  $\phi \sim \phi + 2 \pi$ is the angular coordinate and the boundary is located at $r=\it{const}$. Then, the metric can be written as \cite{Barnich:2014cwa}
\begin{equation}
    \di s^{2}=\mathcal{M}\di u^{2}-2\di u \di r +\mathcal{N}\di\phi \di u+r^{2}\di \phi^{2}\,.
\end{equation}
where $\mathcal{M}$ and $\mathcal{N}$ are two arbitrary functions of $u,\phi$. As was previously discussed, the deformed Maxwell symmetry \eqref{algebra01} can be written in a certain basis as the direct sum $\mathfrak{iso}(2,1)\oplus\mathfrak{so}(2,1)$. Following the same trick considered in \cite{Concha:2018jjj}, we can find  solutions of the present theory by first working in the direct sum basis and then, go to the basis we are interested in. In the direct sum basis, the fields $(\Tilde{\omega}^a,\Tilde{e}^a)$ associated to the $\mathfrak{iso}(2,1)$ generators obey the very well-known GR boundary conditions, and the field $\Tilde{f}^a$ associated to the $\mathfrak{so}(2,1)$ generator can be set as a flat connection.

Furthermore, in the aforementioned direct sum basis, the functions $\mathcal{M}$ and $\mathcal{N}$ are given for the known results in asymptotically flat gravity in three dimensions
\begin{equation}\label{M,N}
    \mathcal{M}=\mathcal{M}(\phi)\,,\qquad \mathcal{N}=\mathcal{J}(\phi)+u \mathcal{M}^{\prime}(\phi)\,.
\end{equation}
 The spacetime line element can be written in terms of the dreibein as $\di s^2 = \eta_{ab} \Tilde{e}^{a} \Tilde{e}^{b}$, where
\begin{equation}
    \eta_{ab} = \begin{pmatrix}
0 & 1 & 0\\
1 & 0 & 0\\
0 & 0 & 1
\end{pmatrix}
\end{equation}
is the Minkowski metric in null coordinate system. Then, the dreibein and the torsionless spin connection one-forms are written as
%\begin{equation}
%\begin{split}
    %&\Tilde{e}^{0}=-\di r+\frac{1}{2}\mathcal{M}\di u+\frac{1}{2}\mathcal{N}\di \phi\,, \qquad \Tilde{e}^{1}= \di u\,, \qquad\Tilde{e}^{2}= r\di\phi \,, \\
  %  &\Tilde{\omega}^{0}=\frac{1}{2}\mathcal{M}\di \phi\,, \qquad\Tilde{\omega}^{1}= \di \phi\,, \qquad \Tilde{\omega}^{2}=0\,.
    %\end{split}
%\end{equation}
\begin{align}
     \Tilde{e}^{0}&=-\di r+\frac{1}{2}\mathcal{M}\di u+\frac{1}{2}\mathcal{N}\di \phi\,, & \Tilde{e}^{1}&= \di u\,, & \Tilde{e}^{2}&= r\di\phi \,, \nonumber \\ 
     \Tilde{\omega}^{0}&=\frac{1}{2}\mathcal{M}\di \phi\,, & \Tilde{\omega}^{1}&= \di \phi\,, & \Tilde{\omega}^{2}&=0\,.
\end{align}
The field $\Tilde{f}^a$ can then be chosen as a Lorentz flat connection, 
\begin{equation}
%\begin{split}
    \Tilde{f}^{0}=\frac{1}{2}\mathcal{L} \di \phi , \qquad\Tilde{f}^{1}= \di \phi, \qquad\Tilde{f}^{2}= 0\,.
 %   \end{split}
\end{equation}
where $\mathcal{L}=\mathcal{L}(\phi)$. In this way, we have found the solutions in the BMS gauge for the fields $(\Tilde{e}^a, \Tilde{\omega}^a, \Tilde{f}^a)$. 

As   mentioned before, we are interested in the basis where the Poincaré-Lorentz symmetry appears as a deformation of the Maxwell algebra. From (\ref{redefine-pp}), it is possible to show that the fields $({e}^a, {\omega}^a, {f}^a)$ are related to those in the direct sum basis $(\Tilde{e}^a, \Tilde{\omega}^a, \Tilde{f}^a)$ as follows:  \begin{equation}
    e^a=\ell\left(\Tilde{f}^a-\Tilde{\omega}^a\right)\,,\qquad \omega^a=\Tilde{\omega}^a\,,\qquad f^a=\ell^{2}\left(\Tilde{f}^a-\Tilde{\omega}^a\right)-\ell\Tilde{e}^a\,.
    \end{equation}
Consequently, the field equations (\ref{EOM}) are solved by the following components of the gauge fields
\begin{align}
    e ^{0} & =\frac{1}{2}\mathcal{P} \di \phi\,,  &
\omega ^{0} & =\frac{1}{2}\mathcal{M}\di \phi\,, & f ^{0} & =\ell \di r+\frac{1}{2}\mathcal{F}\di \phi-\frac{\ell}{2}\mathcal{M}\di u\,,\nonumber \\ \label{ewf}
e ^{1} & =0\,, & \omega ^{1} & =\di \phi\,, & f ^{1} & =-\ell \di u\,, \\ \nonumber
e ^{2} & =0\,, & \omega ^{2} & =0\,, & f^{2} &
=-\ell r \di \phi\,,
\end{align}
where, for later convenience, we have defined the functions $\mathcal{P}=\ell (\mathcal{L}-\mathcal{M})$ and $\mathcal{F}=\ell(\mathcal{P}-\mathcal{N})$.

As discussed we need to ensure  vanishing of the boundary term in the variation of the action when suitable boundary conditions on the fields are imposed. %Now we show that in the space-time with null boundary the action principle is, effectively, \textcolor{blue}{well-posed} \hanote{I think this is a too generic statement and We do not show that.}. 
%As we will see in the next section,
The radial dependence of the connection $A$ can be gauged away by the gauge transformation $A=h^{-1}\di h+h^{-1}ah$, where the asymptotic field $a=a_{u}(u,\phi)\di u+a_{\phi}(u,\phi)\di \phi$ does not depend on $r$ and $h=e^{\ell r M_{0}}$. Then, at the boundary $r=const.\rightarrow \infty$, the on-shell action (\ref{bdy}) takes the form
\begin{align}
   \delta S_{\text{CS}}\big|_{\text{bdy}}&=-\frac{k}{4\pi}\int_{\partial\mathcal{M}} \langle a\delta a\rangle \nonumber\\
&=\frac{k}{4\pi}\int_{\partial\mathcal{M}}\di u \di \phi\,\left[\delta e^{a}_{u}\left(\alpha_{1}\omega_{a\phi}+\left(\frac{\alpha_{1}}{\ell}+\alpha_{2}\right)e_{a\phi}\right)-\delta e^{a}_{\phi}\left(\alpha_{1}\omega_{a u}+\left(\frac{\alpha_{1}}{\ell}+\alpha_{2}\right)e_{a u}\right)\right.  \nonumber \\
     & +\left.\delta \omega^{a}_{u}\left(\alpha_{0}\omega_{a\phi}+\alpha_{1}e_{a\phi}+\alpha_{2}f_{a\phi}\right)-\delta \omega^{a}_{\phi}\left(\alpha_{0}\omega_{a u}+\alpha_{1}e_{a u}+\alpha_{2}f_{a u}\right) \right.\\
     &\left. +  \alpha_2\delta f^{a}_{u}\omega_{a\phi}-\alpha_2\delta f^{a}_{\phi}\omega_{a u} \right] \,.\nonumber
\end{align}
 Furthermore, from (\ref{ewf}) we find the following boundary conditions for the gauge fields
\begin{equation}\label{bdy-condition}
    e^{a}_{u}=0\,, \qquad \omega^{a}_{u}=0\,, \qquad \omega^{a}_{\phi}=-\ell f^{a}_u\,,
\end{equation}
upon the first two,  the variation of the action reduces to
\begin{equation}
    \delta S_{\text{CS}}\big|_{\text{bdy}}=\frac{k \alpha_2}{4\pi}\int_{\partial\mathcal{M}}\di u \di \phi\,\left(\delta \omega^{a}_{\phi}f_{au}-\delta f^{a}_{u} \omega_{a\phi} \right)\,.
\end{equation}
Finally,  applying the last condition $\omega^{a}_{\phi}=-\ell f^{a}_u$, we arrive at
\begin{equation}
    \delta S_{\text{CS}}\big|_{\text{bdy}}=0\,,
\end{equation}
for any value of $\alpha_2$. Thus, in space-time with
%null boundary
 boundary conditions \eqref{bdy-condition}, the action principle is well-posed.

%%%%%%%%%%%%%%%%%%%%%%%%%%%%%%%%%%%%%%%%%%%%%%%%%%%%%%%%%%%%%%%%%%%%%%%%%%%%
%%%%%%%%%%%%%%%%%%%%%%%%%%%%%%%%%%%%%%%%%%%%%%%%%%%%%%%%%%%%%%%%%%%%%%%%%%%%

\section{Asymptotic symmetries}\label{sec:3}
The aim of this section is to find the asymptotic symmetry algebra for the Maxwell CS gravity with torsion which was  previously constructed. To start with, we provide the suitable fall-off conditions for the gauge fields at infinity and the gauge transformations which preserve our boundary conditions. Then, the charge algebra is found using the Regge-Teitelboim method \cite{REGGE1974286}.
\subsection{Boundary conditions}
 Inspired by the results obtained in the previous section, we consider the gauge connection evaluated in the BMS gauge as follows
\begin{equation}
\begin{split}\label{Connection}
       A = &\frac{1}{2}\mathcal{M}\left(u,\phi\right) \di \phi J_0 + \di \phi J_1 + \frac{1}{2}\mathcal{P}\left(u,\phi\right) \di \phi P_0 + \left[\ell \di r+\frac{1}{2}\mathcal{F}\left(u,\phi\right)\di\phi-\frac{\ell}{2}\mathcal{M}\left(u,\phi\right)\di u\right]M_0\\
     & -\ell \di u M_1-r\ell \di\phi M_2\,.
\end{split}
\end{equation}
The radial dependence can be gauged away by an appropriate gauge transformation on the connection
\begin{equation}
    A = h^{-1} \di h +h^{-1} a h\,,
\end{equation}
where the group element is given by $h=e^{\ell r M_0}$. Then, if we use the identity $h^{-1}\di h=\ell \di r M_0$, and the Baker-Campbell-Hausdorff formula, we find
\begin{equation}
   h^{-1} a h = a-\ell r \di\phi M_2\,. 
\end{equation}
Therefore, once we have dropped out the radial dependence from the gauge field $A$, we are left with the asymptotic field $a=a_{u}\di u+a_{\phi}\di \phi$, whose components are given by
\begin{equation}
    \begin{split}
        &a_{u}=-\frac{\ell}{2}\mathcal{M}M_0-\ell M_1\,,\\
        &a_{\phi}=\frac{1}{2}\mathcal{M} J_0+J_1+\frac{1}{2} \mathcal{P} P_0+\frac{1}{2}\mathcal{F}M_0\,,
    \end{split}
\end{equation}
which depend only on time and the angular coordinate. The equations of motion, which are required to hold in the asymptotic region, %leading to the following conditions on the functions $\mathcal{M}$, $\mathcal{P}$ and $\mathcal{F}$:
%\begin{equation}
    %\partial_{u}\mathcal{M}=0\,, \qquad \partial_{u}\mathcal{P}=0\,, \qquad \partial_{u}\mathcal{F}=-\ell \partial_{\phi}\mathcal{M}\,,
%\end{equation}
%which are solved by}\textcolor{blue}{The equations of motion
imply that
\begin{equation}\label{MLN}
   \mathcal{M}=\mathcal{M}\left(\phi\right)\,, \qquad \mathcal{P}=\mathcal{P}\left(\phi\right)\,, \qquad \mathcal{F}=\mathcal{Z}\left(\phi\right)-u \ell \mathcal{M^{\prime}}\left(\phi\right)\,.
\end{equation}

%%%%%%%%%%%%%%%%%%%%%%%%%%%%%%%%%%%%%%%%%%%%%%%%%%%%%%%%%
%%%%%%%%%%%%%%%%%%%%%%%%%%%%%%%%%%%%%%%%%%%%%%%%%%%%%%%%%

\subsection{Residual gauge transformations}
Asymptotic symmetries correspond to residual gauge transformations $\delta_{\Lambda}A = \di{\Lambda}+[A,\Lambda]$ which preserve boundary conditions \eqref{Connection}. We consider the following gauge parameters
%As it is very well-known, the asymptotic symmetry corresponds to residual gauge transformations $\delta_{\Lambda}A = d{\Lambda}+[A,\Lambda]$, preserving our boundary conditions \eqref{Connection}. Let us focus now on finding this set of gauge transformations. We consider the following gauge parameters
\begin{equation}
    \Lambda=h^{-1}\lambda h\,, \qquad \lambda=\lambda^{a}_{(J)}\left(u,\phi\right) J_a
+\lambda^{a}_{(P)}\left(u,\phi\right) P_a+\lambda^{a}_{(M)}\left(u,\phi\right) M_a\,.
    %\lambda=\chi^{a}\left(u,\phi\right) J_a +\varepsilon^{a}\left(u,\phi\right) P_a+\gamma^{a}\left(u,\phi\right) M_a\,.
\end{equation}
Then, gauge transformations of the connection $A$ with gauge parameter $\Lambda$, lead to $r$-independent gauge transformations of the connection $a$ with gauge parameter $\lambda$, i.e.
\begin{equation}
    \delta_{\Lambda}a \equiv \delta_{\lambda}a=\di\lambda+[a,\lambda] \, .
\end{equation}
The gauge transformations that preserve the boundary conditions \eqref{Connection} with \eqref{MLN} for $\lambda_{(J)}^{a}$ and $\lambda_{(P)}^{a}$ are given by
\begin{align}\label{deltaa-1}
    \lambda_{(J)}^{0} &=\frac{\mathcal{M}}{2}\,\varepsilon-\varepsilon^{\prime \prime }\,,    &  \lambda_{(P)}^{0} & =\frac{1}{2}\left[\mathcal{P}\left(\frac{\chi}{\ell}+ \varepsilon\right)+ \mathcal{M} \chi\right] -\chi^{\prime \prime }\,, \nonumber \\ 
    \lambda_{(J)}^{1} & =\varepsilon\,,   & \lambda_{(P)} ^{1} & =\chi \,, \\
    \lambda_{(J)}^{2} & =-\varepsilon^{\prime }\,, & \lambda_{(P)}^{2} & =-\chi^{\prime} \nonumber\,,
\end{align}
while for $\lambda_{(M)}^{a}$ we get
\begin{align}\label{deltaa-2}
    \lambda_{(M)}^{0}& =\frac{\mathcal{M}}{2}\gamma+\frac{\mathcal{P}}{2} \chi+\frac{ \mathcal{F}}{2}\varepsilon-\frac{\ell \mathcal{M}}{2}u \varepsilon^{\prime }+u\ell  \varepsilon^{\prime\prime\prime }-\gamma^{\prime\prime}\,, \nonumber\\
    \lambda_{(M)}^{1} & =\gamma-u \ell \varepsilon^{\prime }\,, \\
    \lambda_{(M)}^{2} & =-\gamma^{\prime }+u \ell \varepsilon^{\prime \prime } \nonumber\,.
    \end{align}
where $\varepsilon$, $\chi$ and $\gamma$ are three arbitrary, periodic functions of the angular coordinate $\phi$. Under the given gauge transformation, the dynamical fields transform as
\begin{eqnarray}
\delta_\lambda \mathcal{M} &=&\mathcal{M}^{\prime }\varepsilon+2\mathcal{M}\varepsilon^{\prime
}-2\varepsilon{}^{\prime \prime \prime }\,,  \notag \\
\delta_\lambda \mathcal{P} &=&\mathcal{P}^{\prime }\left(\frac{\chi}{\ell}+\varepsilon\right)+2\mathcal{P}\left(\varepsilon^{\prime}+\frac{\chi^{\prime
}}{\ell}\right)+\mathcal{M}^{\prime }\chi+2\mathcal{M}\chi^{\prime}-2\chi^{\prime \prime \prime }\,,  \label{translaw} \\
\delta_\lambda \mathcal{Z} &=&\mathcal{Z}^{\prime}\varepsilon+2\mathcal{Z}\varepsilon^{\prime}+\mathcal{M}^{\prime}\gamma+2\mathcal{M}\gamma^{\prime}+\mathcal{P}^{\prime}\chi+2\mathcal{P}\chi^{\prime}-2\gamma^{\prime \prime \prime}\,.  \notag
\end{eqnarray}

\subsection{Canonical surface charges and asymptotic symmetry algebra}
 Asymptotic symmetries of the Maxwell gravity theory with non-vanishing torsion can be found in the canonical approach \cite{REGGE1974286}. In particular, in the case of a three-dimensional Chern-Simons theory, the variation of the canonical generators is given by \cite{Banados:1998gg,Banados:1994tn}
\begin{equation}
    \delta Q[\lambda]=\int d\phi \langle \lambda \delta a_\phi\rangle\,.
\end{equation}
Therefore one can show that surface charge variation associated with \eqref{deltaa-1} and \eqref{deltaa-2} is
%Then, \textcolor{blue}{using (\ref{deltaa}) and} the invariant tensor (\ref{invtensor}), it is possible to show that the above expression is linear in the variation of the functions $\mathcal{M}$, $\mathcal{J}$ and $\mathcal{L}$, so that it can be integrated leading to the following expression
\begin{equation}\label{charge-variation-01}
    \delta Q (\varepsilon,\chi, \gamma )= \int_0^{2\pi} \di \phi \left( \varepsilon \delta \mathbf{J}+ \chi \delta \mathbf{P} + \gamma \delta \mathbf{M} \right)
\end{equation}
with
\begin{subequations}
\begin{align}
   \mathbf{J} =& \frac{k}{4\pi}\left(\alpha_2\mathcal{Z}+\alpha_0\mathcal{M}+\alpha_1\mathcal{P}\right)\, , \\
   \mathbf{P}= & \frac{k}{4\pi}\left[\left(\frac{\alpha_{1}}{\ell}+ \alpha_{2}\right)\mathcal{P}+ \alpha_{1}\mathcal{M}\right]\, , \\
   \mathbf{M}= & \frac{k}{4\pi} \alpha_{2} \mathcal{M} \, .
\end{align}
\end{subequations}
One can take $\varepsilon$, $\chi$ and $\gamma$ to be state-independent and then the charge variation \eqref{charge-variation-01} is integrable on the phase space.
%\begin{align}
%    Q\left[Y,R,T\right]& =\frac{k}{4\pi}\int d\phi \left[Y \left(\alpha_2\mathcal{Z}+\alpha_0\mathcal{M}+\alpha_1\mathcal{P}\right)\right. \nonumber \\ 
%   & +\left.  R \left(\left(\frac{\alpha_{1}}{\ell}+ \alpha_{2}\right)\mathcal{P}+ \alpha_{1}\mathcal{M}\right)+\alpha_{2}T\mathcal{M}\right]\,.\label{QYRT}
%\end{align}
%Now, from (\ref{QYRT}) we can see that there are three independent terms. We define the asymptotic charges corresponding to the aforesaid terms as follows
There are three independent surface charges,
\begin{equation}
    J(\varepsilon) = Q(\varepsilon,0,0) \, , \qquad P(\chi) = Q(0,\chi,0) \, , \qquad M(\gamma) = Q(0,0,\gamma) \, , \qquad 
\end{equation}
associated with three independent symmetry generators $\varepsilon$, $\chi$ and $\gamma$.
 %\begin{eqnarray}
 %j[Y] &=&\frac{k}{4\pi}\int d\phi\, Y\left(\alpha_2\mathcal{Z}+\alpha_0\mathcal{M}+\alpha_1\mathcal{P}\right)\,,  \nonumber \\
 %p[R]&=&\frac{k}{4\pi}\int d\phi\, R\left[\left(\frac{\alpha_1}{\ell}+\alpha_2\right)\mathcal{P}+\alpha_1\mathcal{M} %\right]\,,\label{jpz}\\ \nonumber
 %\textsc{m}[T]&=&\frac{k}{4\pi}\int d\phi\, \alpha_{2}T\mathcal{M}\,.
 %\end{eqnarray}
It is shown that the algebra among surface charges is given by \cite{REGGE1974286,brown1986poisson,compere2019advanced}
\begin{equation}
    \{ Q(\Lambda_{1}), Q(\Lambda_{2}) \} = Q([\Lambda_{1},\Lambda_{2}])+ \mathcal{C}(\Lambda_{1},\Lambda_{2})
\end{equation}
where Dirac bracket is defined as $\{ Q(\Lambda_{1}), Q(\Lambda_{2}) \}:= \delta_{\Lambda_{2}}Q(\Lambda_1)$ and $\mathcal{C}(\Lambda_{1},\Lambda_{2})$ is central extension term. Therefore, using the transformation laws \eqref{translaw}, one can show that the algebra of charges \eqref{charge-variation-01} is
\begin{eqnarray}
\left\lbrace J(\varepsilon_1),J(\varepsilon_2)\right\rbrace&=&J([\varepsilon_1,\varepsilon_2])-\frac{k\alpha_0}{2\pi}\int \di \phi \, \varepsilon_1 \varepsilon^{\prime\prime\prime}_2\,, \nonumber\\
\left\lbrace J(\varepsilon),P(\chi)\right\rbrace&=&P([\varepsilon,\chi])-\frac{k\alpha_1}{2\pi} \int \di \phi
\, \varepsilon \chi^{\prime\prime\prime}\,, \nonumber\\
\left\lbrace J(\varepsilon),M(\gamma)\right\rbrace&=&M([\varepsilon,\gamma])-\frac{k\alpha_2}{2\pi} \int \di \phi
\, \varepsilon \gamma^{\prime\prime\prime}\,,\\
\left\lbrace P(\chi_1),P(\chi_2)\right\rbrace&=&M([\chi_1,\chi_2])+\frac{1}{\ell} P([\chi_1,\chi_2])-\frac{k}{2\pi}\left(\frac{\alpha_1}{\ell}+\alpha_2\right)\int \di \phi \chi_1 \chi^{\prime\prime\prime}_2\,, \nonumber\\
\left\lbrace P(\chi),M(\gamma)\right\rbrace&=&0\,, \nonumber\\
\left\lbrace M(\gamma_1),M(\gamma_2)\right\rbrace&=&0\,,  \nonumber
\end{eqnarray}

\begin{comment}
The Poisson algebra can be evaluated from the variation under gauge transformations \cite{REGGE1974286}
\begin{equation}
    \delta_{\Lambda_{2}}Q\left[\Lambda_1\right]=\left\lbrace Q[\Lambda_{1}],Q[\Lambda_{2}]\right\rbrace\,.
\end{equation}
Then, using the previous equation and the transformation laws (\ref{translaw}), the Poisson brackets of the defined charges (\ref{jpz}) can be evaluated leading to
\begin{eqnarray}
\left\lbrace j[Y_1],j[Y_2]\right\rbrace&=&j[[Y_1,Y_2]]-\frac{k\alpha_0}{2\pi}\int d\phi Y_1Y^{\prime\prime\prime}_2\,, \nonumber\\
\left\lbrace j[Y],p[R]\right\rbrace&=&p[[Y,R]]-\frac{k\alpha_1}{2\pi} \int d\phi
Y R^{\prime\prime\prime}\,, \nonumber\\
\left\lbrace j[Y],\textsc{m}[T]\right\rbrace&=&\textsc{m}[[Y,T]]-\frac{k\alpha_2}{2\pi} \int d\phi
Y T^{\prime\prime\prime}\,,\\
\left\lbrace p[R_1],p[R_2]\right\rbrace&=&\textsc{m}[[R_1,R_2]]+\frac{1}{\ell} p[[R_1,R_2]]-\frac{k}{2\pi}\left(\frac{\alpha_1}{\ell}+\alpha_2\right)\int d\phi R_1 R^{\prime\prime\prime}_2\,, \nonumber\\
\left\lbrace p[R],\textsc{m}[T]\right\rbrace&=&0\,, \nonumber\\
\left\lbrace \textsc{m}[T_1],\textsc{m}[T_2]\right\rbrace&=&0\,,  \nonumber
\end{eqnarray}
\end{comment}
where $[x,y]=x y^{\prime}-y x^{\prime}$.
%where the Lie bracket is denoted as $[x,y]=x y^{\prime}-y x^{\prime}$.
The resulting algebra corresponds to an infinite-dimensional lift of the deformed Maxwell algebra.  Now we can express the above algebra in Fourier modes,
\begin{equation}
    J_n := J(e^{in\phi})\,, \qquad P_n := P(e^{in\phi})\,, \qquad M_n := M(e^{in\phi})\,,
\end{equation}
%\begin{equation}
%    J_{n}=j[e^{in\phi}]\,, \qquad P_{n}=p[e^{in\phi}]\,, \qquad M_{n}=\textsc{m}[e^{in\phi}]\,, \qquad m\in\mathbb{Z}\,,
%\end{equation}
which give rise to the following centrally extended algebra
\begin{eqnarray}
i\left\lbrace J_m,J_n\right\rbrace &=& \left(m-n\right)J_{m+n}+\frac{c_1}{12}m^3\delta_{m+n,0}\,,\nonumber\\
i\left\lbrace J_m,P_n\right\rbrace &=& \left(m-n\right)P_{m+n}+\frac{c_2}{12}m^3\delta_{m+n,0}\,,\nonumber\\
i\left\lbrace J_m,M_n\right\rbrace &=& \left(m-n\right)M_{m+n}+\frac{c_3}{12}m^3\delta_{m+n,0}\,,\nonumber\\
i\left\lbrace P_m,P_n\right\rbrace &=&
\left(m-n\right)M_{m+n} +\frac{1}{\ell}\left(m-n\right)P_{m+n}+\frac{1}{12}\left(\frac{c_2}{\ell} +c_3\right)m^3\delta_{m+n,0}\,,\label{inf-algebra}\\
i\left\lbrace P_m,M_n\right\rbrace &=&0\,,\nonumber\\
i\left\lbrace M_m,M_n\right\rbrace &=&0\,.\nonumber
\end{eqnarray}
%\begin{eqnarray}
%i\left\lbrace J_m,J_n\right\rbrace &=& \left(m-n\right)J_{m+n}+\frac{c_1}{12}m^3\delta_{m+n,0}\,,\nonumber\\
%i\left\lbrace J_m,P_n\right\rbrace &=& \left(m-n\right)P_{m+n}+\frac{c_2}{12}m^3\delta_{m+n,0}\,,\nonumber\\
%i\left\lbrace J_m,M_n\right\rbrace &=& \left(m-n\right)M_{m+n}+\frac{c_3}{12}m^3\delta_{m+n,0}\,,\nonumber\\
%i\left\lbrace P_m,P_n\right\rbrace &=&
%\left(m-n\right)M_{m+n} +\frac{1}{\ell}\left(m-n\right)P_{m+n}+\frac{1}{12}\left(\frac{c_2}{\ell} +c_3\right)m^3\delta_{m+n,0}\,,\label{inf-algebra}\\
%i\left\lbrace P_m,M_n\right\rbrace &=&0\,,\nonumber\\
%i\left\lbrace M_m,M_n\right\rbrace &=&0\,.\nonumber
%\end{eqnarray}
%Here, we have defined the three central charges $c_{1}$, $c_{2}$ and $c_{3}$ in terms of the CS level $k$ and the arbitrary constant appearing in the invariant tensor (\ref{invtensor}) as follows
The central charges $c_{1}$, $c_{2}$ and $c_{3}$ are related to the CS level $k$ and the arbitrary constant appearing in the invariant tensor \eqref{invtensor} as
\begin{equation}
    c_{i}=12 k\alpha_{i-1}\,.
\end{equation}

%%%%%%%%%%%%%%%%%%%%%%%%%%%%%%%%%%%%%%%%%%%%%%%%%%%%%%%%%%%%%%
\subsection{Change of basis}
%%%%%%%%%%%%%%%%%%%%%%%%%%%%%%%%%%%%%%%%%%%%%%%%%%%%%%%%%%%%%%

The infinite-dimensional algebra \eqref{inf-algebra} can be seen as the infinite-dimensional enhancement of the deformed Maxwell algebra \eqref{algebra01}. In particular, in the flat limit $\ell\rightarrow\infty$ we recover the asymptotic symmetry of the three-dimensional Maxwell CS gravity theory introduced in \cite{Concha:2018zeb}. Interestingly, as was shown in \cite{Concha:2019eip}, the algebra \eqref{inf-algebra} is isomorphic to the $\widehat{\mathfrak{bms}}_3\oplus \mathfrak{vir}$ algebra.
To observe this at the level of charge algebra, we use the change of basis proposed in \cite{Grumiller:2019fmp}.

Suppose $\varepsilon$, $\chi$ and $\gamma$ are now state-dependent, i.e. they are functions of dynamical fields. We require that charge variation be integrable which leads to
\begin{equation}
    \varepsilon = \frac{\delta \mathcal{G}}{\delta \mathbf{J}}\, , \qquad \chi = \frac{\delta \mathcal{G}}{\delta \mathbf{P}}\, , \qquad \gamma = \frac{\delta \mathcal{G}}{\delta \mathbf{M}}\, ,
\end{equation}
for some functional
\begin{equation}
    \mathcal{G}[\mathbf{J},\mathbf{P},\mathbf{M}]= \int_0^{2\pi} \di \phi \, \mathbf{G}(\mathbf{J},\mathbf{P},\mathbf{M})\, .
\end{equation}
By choosing
\begin{equation}
    \mathbf{G}= \tilde{\varepsilon} \left( \mathbf{J}- \ell \mathbf{P}- \ell^2 \mathbf{M} \right) +\tilde{\chi} \left( \ell \mathbf{P}+ \ell^2 \mathbf{M}\right)- \ell \tilde{\gamma} \,\mathbf{M}
\end{equation}
where $\tilde{\varepsilon},\tilde{\chi}, \tilde{\gamma}$ are state-independent functions, one can show that the charge variation \eqref{charge-variation-01} can be written as
\begin{equation}\label{charge-variation-02}
    \delta Q (\tilde{\varepsilon},\tilde{\chi}, \tilde{\gamma} )= \int_0^{2\pi} \di \phi \left( \tilde{\varepsilon} \delta \mathbf{L}+ \tilde{\chi} \delta \mathbf{S} + \tilde{\gamma} \delta \mathbf{T} \right)
\end{equation}
with
\begin{subequations}
\begin{align}
   \mathbf{L} =& \mathbf{J}- \ell \mathbf{P}- \ell^2 \mathbf{M}\, , \\
   \mathbf{S}= & \ell \mathbf{P}+ \ell^2 \mathbf{M}\, , \\
   \mathbf{T}= &  - \ell \,\mathbf{M}\, .
\end{align}
\end{subequations}
Now, by introducing Fourier modes
\begin{equation}
   L_m := Q (\tilde{\varepsilon}=e^{in\phi},0,0 ),\qquad S_m:= Q (0,\tilde{\chi}=e^{in\phi},0 ) ,\qquad T_{m}:=Q (0,0, \tilde{\gamma}=e^{in\phi} ),
\end{equation}
one finds the direct sum of the $\widehat{\mathfrak{bms}}_3$ and Virasoro algebra:
\begin{eqnarray}\label{bms3+vir}
i\left\lbrace L_m,L_n\right\rbrace &=& \left(m-n\right)L_{m+n}+\frac{c_{LL}}{12}m^3\delta_{m+n,0}\,,\nonumber\\
i\left\lbrace L_m,T_n\right\rbrace &=& \left(m-n\right)T_{m+n}+\frac{c_{LT}}{12}m^3\delta_{m+n,0}\,,\label{bmsVir}\\
i\left\lbrace S_m,S_n\right\rbrace &=& \left(m-n\right)S_{m+n}+\frac{c_{SS}}{12}m^3\delta_{m+n,0}\,,\nonumber
\end{eqnarray} where the central charges are related to those appearing in \eqref{inf-algebra} as
%\begin{equation}
  %  c_{1}\equiv c_{LL}+c_{SS}\,, \qquad c_{2}\equiv c_{LT}+\frac{1}{\ell}c_{SS}\,, \qquad c_{3}\equiv -\frac{1}{\ell}c_{LT}\,.
%\end{equation}
\begin{equation}
    c_{LL}\equiv c_{1}-\ell c_{2}-\ell^2 c_{3}\,, \qquad c_{SS}\equiv \ell c_{2}+\ell^2 c_{3}\,, \qquad c_{LT}\equiv -\ell c_{3}\,.
\end{equation} 
The central charges $\left(c_{LL},c_{LT},c_{SS}\right)$ of the $\widehat{\mathfrak{bms}}_3\oplus\mathfrak{vir}$ algebra are related to the three independent terms of the CS gravity action for the $\mathfrak{iso}(2,1)\oplus\mathfrak{so}(2,1)$ algebra.

\begin{comment}
Indeed, by considering the following redefinition of generators,
\begin{equation}
    J_{m}\equiv L_{m}+S_{m}\,, \qquad P_{m}\equiv T_{m}+\frac{1}{\ell}S_{m}\,, \qquad M_{m}\equiv -\frac{1}{\ell}T_{m}\,,
\end{equation}
one finds the direct sum of the $\widehat{\mathfrak{bms}}_3$ and Virasoro algebra:
\begin{eqnarray}
i\left\lbrace L_m,L_n\right\rbrace &=& \left(m-n\right)L_{m+n}+\frac{c_{LL}}{12}m^3\delta_{m+n,0}\,,\nonumber\\
i\left\lbrace L_m,T_n\right\rbrace &=& \left(m-n\right)T_{m+n}+\frac{c_{LT}}{12}m^3\delta_{m+n,0}\,,\label{bmsVir}\\
i\left\lbrace S_m,S_n\right\rbrace &=& \left(m-n\right)S_{m+n}+\frac{c_{SS}}{12}m^3\delta_{m+n,0}\,,\nonumber
\end{eqnarray} where the central charges are related to those appearing in \eqref{inf-algebra} as
\begin{equation}
    c_{1}\equiv c_{LL}+c_{SS}\,, \qquad c_{2}\equiv c_{LT}+\frac{1}{\ell}c_{SS}\,, \qquad c_{3}\equiv -\frac{1}{\ell}c_{LT}\,.
\end{equation}
Here, the central charges $\left(c_{LL},c_{LT},c_{SS}\right)$ of the $\widehat{\mathfrak{bms}}_3\oplus\mathfrak{vir}$ algebra are related to the three independent terms of the CS gravity action for the $\mathfrak{iso}(2,1)\oplus\mathfrak{so}(2,1)$ algebra.
\end{comment}

Let us note that the algebra \eqref{inf-algebra} can be obtained alternatively as a central extension of the deformed \Max \cite{Concha:2019eip} and deformed $\mathfrak{bms}_{4}$ \cite{Safari:2019zmc}. It is also worth  pointing out that by ignoring the generators $T_{m}$ and central charge $c_{LT}$ in \eqref{bms3+vir}, which is equivalent to ignoring the Maxwell generators $M_{m}$ and central charge $c_{3}$ in \eqref{inf-algebra}, one obtains two copies of Virasoro algebra, which is the asymptotic symmetry of the AdS$_{3}$ CS gravity (or Teleparallel theory \cite{Blagojevic:2003uc}) with two central charges as 
\begin{equation}
    c_{LL}\equiv c_{1}-\ell c_{2}\,, \qquad c_{SS}\equiv \ell c_{2}\,.
\end{equation}

\begin{comment}
In particular, two copies of the Virasoro algebra appear by identifying
\begin{equation}
J_{m}\equiv L_{m}+\bar{L}_{m}\,, \qquad P_{m}\equiv \frac{1}{\ell}L_{m}\,, \qquad  c_{1}\equiv c+\bar{c}\,, \qquad c_{2}\equiv \frac{1}{\ell}c\,.
\end{equation}
where $L_{m}$ and $\bar{L}_{m}$ are the Virasoro generators satisfying
\begin{eqnarray}
i\left\lbrace L_m,L_n\right\rbrace &=& \left(m-n\right)L_{m+n}+\frac{c}{12}m^3\delta_{m+n,0}\,,\nonumber\\
i\left\lbrace \bar{L}_m,\bar{L}_n\right\rbrace &=& \left(m-n\right)\bar{L}_{m+n}+\frac{\bar{c}}{12}m^3\delta_{m+n,0}\,.\label{Vir}
\end{eqnarray}
\end{comment}

%Naturally, when we consider $M_{m}=0$, $c_{3}=0$ and the flat limit $\ell\rightarrow\infty$ in \eqref{inf-algebra}, the asymptotic structure corresponds to the $\widehat{\mathfrak{bms}}_3$ algebra.
%\hnote{The last sentence is a bit confusing since one can obtain the $\widehat{\mathfrak{bms}}_3$ algebra as sub algebra of \eqref{bmsVir} by another limit. I think this is not necessary to mention it here. }
%%%%%%%%%%%%%%%%%%%%%%%%%%%%%%%%%%%%%%%%%%%%%%%%%%%%%%%%%%%%%%%%%%%%%%%%%%%%
%%%%%%%%%%%%%%%%%%%%%%%%%%%%%%%%%%%%%%%%%%%%%%%%%%%%%%%%%%%%%%%%%%%%%%%%%%%%

\section{Discussion and outlook}\label{sec-discussions}

In this work, we have studied a CS gravity theory with
%based on
a deformed Maxwell algebra as a gauge group,  which allows us to introduce a non-vanishing torsion to the Maxwell CS gravity. In particular, the CS action can be seen as a Maxwell generalization of a particular case of the MB gravity \cite{Mielke:1991nn} whose field equations correspond to those of the Maxwell gravity theory but in presence of a non-vanishing torsion. Motivated by the fact that the deformed Maxwell is isomorphic to the $\mathfrak{iso}(2,1)\oplus\mathfrak{so}(2,1)$ algebra, we considered asymptotically flat geometries and discuss the BMS-like solution. We then explored the implications of the deformed Maxwell algebra \eqref{algebra01} at the level of the asymptotic symmetry. In particular, we have shown that the asymptotic symmetry for the Maxwell algebra with torsion is described by an infinite-enhancement of the deformed Maxwell algebra which can be written as the $\widehat{\mathfrak{bms}}_3\oplus\mathfrak{vir}$ algebra with three independent central charges.

It would be interesting to go further in the study of the solutions of the present theory. In particular, one expects  the solutions to describe a Maxwell version of the so-called Teleparallel theory in which the cosmological constant is a source for the torsion. As was shown in \cite{Blagojevic:2003uc,Blagojevic:2003wn}, both GR with cosmological constant and three-dimensional Teleparallel gravity have the same asymptotic symmetry and dynamics. Furthermore, three-dimensional gravity with torsion possesses BTZ \cite{Banados:1992wn} black hole solution \cite{Garcia:2003nm,Mielke:2003xx,Blagojevic:2003vn} whose thermodynamics properties have been discussed in \cite{Blagojevic:2006jk}.
Then, it seems natural to expect to find a BTZ type solution for our theory. 
%The analysis of the conserved charges \hanote{I think we did this part.} and 
The approach to this last point and thermodynamics remains as an interesting open issue to explore. In particular, one can study the effects of the present deformation in the vacuum energy and vacuum angular momentum of the stationary configuration and their relations to the GR and Maxwell ones.

Although the three-dimensional Maxwell algebra can be deformed into $\mathfrak{so}(2,2)\oplus\mathfrak{so}(2,1)$ and  $\mathfrak{iso}(2,1)\oplus\mathfrak{so}(2,1)$ algebras, the former can be obtained as a deformation of the latter. The same argument is true for their corresponding infinite enhancements, meaning that the $\mathfrak{bms}_3\oplus\mathfrak{witt}$ algebra can be deformed into the $\mathfrak{witt}\oplus\mathfrak{witt}\oplus\mathfrak{witt}$ algebra. Since both infinite-dimensional algebras are obtained as asymptotic symmetry algebras of different three-dimensional CS gravity theories, it might be interesting to consider the deformation relation between these theories at the level of their solutions, phase spaces and physical quantities. It should be pointed out that the $\mathfrak{bms}_3\oplus\mathfrak{witt}$ algebra can also be deformed into another algebra which is known as $W(a,b)\oplus\mathfrak{witt}$, as was proved in \cite{Parsa:2018kys}. Regarding the recent work \cite{Grumiller:2019fmp}, where it was shown that the $W(0,b)$ algebra can be obtained as near horizon symmetries of three-dimensional black holes, a natural question that one can ask is what is the interpretation of the $W(a,b)\oplus\mathfrak{witt}$ algebra in the context of the new CS theory. It would be worthwhile to explore the answer.

Another aspect that deserves to be explored is supersymmetric  extension of our analysis and results. Supersymmetric extensions of three-dimensional gravity with torsion has been constructed in \cite{Giacomini:2006dr,Cvetkovic:2007sr}. On the other hand, the three-dimensional CS supergravity invariant under the Maxwell algebra has been recently presented in \cite{Concha:2015woa,Concha:2018jxx,Concha:2019icz,Concha:2019mxx}. However, to our knowledge, a  supersymmetric Maxwell CS gravity with torsion has not been discussed yet. It would be then interesting to explore diverse supersymmetric extensions of the deformed Maxwell algebra and study which superalgebra is a good candidate to construct a well-defined supersymmetric gravity with torsion in three spacetime dimensions. The study of supersymmetric extensions of CS gravity with torsion and their asymptotic structures could bring a better understanding of the role of torsion.

%%%%%%%%%%%%%%%%%%%%%%%%%%%%%%%%%%%%%%%%%%%%%%%%%%%%%%%%%%%%%%%%%%%%%%%%%%%%%%%%%%%%%%%%%%%%%%%%%%%%%%%%%%
\section*{Acknowledgement}
 The authors would like to thank N. Merino and M. M. Sheikh-Jabbari for the correspondence and comments. This work was funded by the National Agency for Research and Development (ANID) CONICYT - PAI grant No. 77190078 (P.C.) and FONDECYT Project No. 3170438 (E.R.). H.A. and H.R.S. acknowledge the partial support of Iranian NSF under grant No. 950124. H.A. acknowledges the support by the Saramadan grant No. ISEF/M/98204. H.R.S. wishes to thank to M. Henneaux and S. Detournay for their kind hospitality at the Physique Théorique et Mathématique department of the Université Libre de Bruxelles (ULB) which part of this work was done and acknowledges the support by F.R.S.-FNRS fellowship "Bourse de séjour scientifique IN". P.C. would like to thank to the Dirección de Investigación and Vice-rectoria de Investigación of the Universidad Católica de la Santísima Concepción, Chile, for their constant support.

%-------------------------------------------------------------------------------------------------------------------------------------
%-------------------------------------------------------------------------------------------------------------------------------------

%%%%%%%%%%%%%%%%%%%%%%%%%%%%%%%%%%%%%%%%%%%%%%%%%%%%%%%%%%%%%%%%%%%%%%%%%%%%%%%%%%%%%%%%%%%%%%%%%%%%%%%

\bibliographystyle{fullsort.bst}
 
\providecommand{\href}[2]{#2}\begingroup\raggedright\endgroup

\end{document}